\documentstyle[12pt]{article}
\title{A Supermassive Black Hole or a Compact Object Without Events Horizon ?}
\author{L.V.Verozub \& E.Yu.Bannikova\\
Kharkov State University, Kharkov 310077,\\
Kharkov, Ukraine, \\E-mail:verozub@gravit.kharkov.ua}
\date{}
\begin{document}
\maketitle
\begin{abstract}
Previously it was shown that gravitation theory allows the existance of
supermassive stable compact configurations of  the degenerated electronic
gas ( L.V.Verozub, Astr. Nacr. 317 (1996) 107 ) without events horizon.
In the present paper the simplest model of such  kind of  objects in gas
environment  has been considered.
It is shown that at the spherically symmetric accretion onto
the object the luminosity is about $10^{37} erg/s$ for the mass accretion
rate of the order of $\stackrel{\cdot}{M}=10^{-6} M_{\odot}/year$ .
The vawelength of the radiation maximum is  about $ 400\div 500
\stackrel{\circ}{A}$. There is an  ionization zone
around the central objects with the radius about  $10^{-3}pc$ .

\end{abstract}

\section{Introduction}
The analysis of the observation data gives evidence for the existance
of a massive
( about $2.5\cdot10^{6} M_{\odot}$ ) compact object in the Galactic Center
\cite{Eckart}.
The observation data do not allow to make a definite conclusion about the
nature of the object. For this reason it is identified , as a rule, with a
supermassive black hole. Another possibility is considered in the present
paper.

The  gravitation equations whose spherically symmetric solution have no
and physical singularity in flat space-time from
the viewpoint of a remote observer where proposed in  the paper
\cite{Verozub1} . According to the equations the events horizon
is absent in the spherically symmetric solution. The radial component of the
gravity force $F$
affecting a test particle with mass $m$  in the spherical coordinate
system in flat space-time is given by
\begin{equation}
F=-m\left[ c^{2}B_{00}^{1} + (2 B_{01}^{0} - B_{11}^{1})v^{2}]\right] .
\label{gravityforce}
\end{equation}
  Here $B_{00}^{1}$, $B_{01}^{0}$ and $B_{11}^{1}$ are the nonzero components
 of the strength tensor $B_{\alpha \beta }^{\gamma }$ of gravity field in
flat space-time :\\
\begin{eqnarray}
B_{00}^{1}=\frac{1}{2}\frac{\alpha f' f^{4} (1-\alpha /f)}{f^2 r^{4}},\\
B_{01}^{0}=2/r,\\
B_{11}^{1}=\frac{1}{2} \frac{\alpha  f'}{f^{2} (1-\alpha /f)},\\
f=(\alpha ^{3}+r^{3}) ,
\end{eqnarray}
$\alpha =2GM/c^{2}$ is the Schwarzshild radius and
$v$ is the radial component of the particle velocity.

Fig 1 shows the value of $|F|$ for the particle at rest as a function of
$\overline{r}=r/\alpha $ .

It is shown in the paper \cite{Verozub2} that in the above theory there
can exist equilibrium stable configurations of the degenerated electronic gas
with
masses up to $10^{9} M_{\odot}$ or more than that. This  kind of
objects there can exist in  the Galactic Center.

Fortunately, there are some consequences available for observations
 which can help us to identify the objects with one of the proposed
hypotheses.

\section{Peculiarity of Accretion Onto the Massive Objects Without
Events Horizon}

Consider the object without the events horizon with the mass
$2.5\cdot M_{\odot}$
in the center of a spherical symmetric gas medium . Let us
assume that the
gas density is sufficient to describe the gas motion by using the
hydrodynamics equations and  the state equation  is
$P=K\rho ^{\gamma }$ where $P$ is the gas pressure, $\rho $ is the density,
$K$ and $\gamma $ are the constants. (For numerical estimates we assume
in this paper that $\gamma =4/3$).

The following equations are used here for describing  the system
\cite{Shapiro}.

1. The integral of the continuity equation
\begin{equation}
4\pi r^2 v \rho =\stackrel{\cdot}{M}
\label{eqncontinuity}
\end{equation}
where $ v$ is the radial gas velocity at the distance $r$ from the center,
$\stackrel{\cdot}{M}$ is the mass rate of accretion .\\
2. The adiabatic relasionship  between the sound velocity $a$ and the
density $\rho $
\begin{equation}
\rho =Const \ a^{2/(\gamma -1)}
\label{rho_velosity}
\end{equation}
3. The Euler equation
\begin{equation}
u u' + a^{2} \frac{\rho '}\rho  + F/m=0
\label{Euler}
\end{equation}
where $F$ is given by eq. (\ref{gravityforce})

The velocity of the gas falling from infinity to the center reaches the
sound velocity $a$ at the distance $r_{s}$ which is defined by the
equations
\begin{eqnarray}
2\frac{a_{s}^{2}}{r_{s}} = F(r_{s},a_{s})\\ \nonumber
r_{s}^2 a^{\frac{\gamma +1}{\gamma -1}} = \frac{\stackrel{\cdot}{M}}{4 \pi Q}
\label{eqnsforrsandas}
\end{eqnarray}
where $a_{s}$ is the value of $a$ at $r=r_{s}$ ,

$$Q=\frac{\rho _{\infty}}{a_{\infty}^{2/(\gamma -1)}} , $$

$v_{\infty}$ and $\rho _{\infty}$ are the velocity and density at infinity.
In contrast to the Newtonian gravity law equations (\ref{eqnsforrsandas})
have two solutions. At  $v_{\infty}=10^{7} cm/s $ , at the density of
the particles number $n=10^{2} cm^{-3}$ and at
$\stackrel{\cdot}{M}=10^{-6} M_{\odot}/year $ , the numerical
solution of eqs. (\ref{eqnsforrsandas}) yields
\begin{center}
\begin{tabular}{l r}
$ r_{1s} = 0.83 \cdot 10^{18}cm$  & $r_{2s} = 1.4 \cdot 10^{11}cm $ \\
$ a_{1s} = 1.40 \cdot 10^{7}cm $  & $ a_{2s} = 1.1 \cdot 10^{9}cm $ \\
\end{tabular}
\end{center}

The reason of the second solution is that the gas velocity of a particle
falling from infinity increases up to the distances of the order of the
Schwarzschild radius and after that decreases according to the perculiarity
of the gravity force.

We find the function $v(r)$ as the result of the numerical solution of the
following equation resulting from eqs. (\ref{eqncontinuity}),
(\ref{rho_velosity}) and (\ref{Euler})
\begin{equation}
v' - \frac{2K}{D} r^{1-2 \gamma } v^{1-\gamma } + \frac{F}{D} = 0
\label{eqfor_v}
\end{equation}
where
$$D=v-K r^{-2(\gamma -1)} v^{-\gamma },$$
$$ K=a^{2}_{\infty} (A/ \rho _{\infty})^{\gamma -1} $$
and
$$A= \stackrel{\cdot}{M}/4\pi $$.

Fig. 2 shows $v$ as the function of $r$ from $r=r_{2s}$
to the surface of the central object. It is the result of the numerical
solution  of eq.  (\ref{Euler}) with the help (\ref{eqncontinuity}) and
(\ref{rho_velosity}). The radius $R$ of the central  object has been  found by
the numerical solution of the equation of the hydrodynamical equilibrium
and is equal to $0.04 \alpha $ , or $10^{11} cm$.

\section{Luminosity}

The Eddington's limit of the luminosity near the surface of the central
object is given by
\begin{equation}
L = \frac{4 \pi  c}{ \sigma } GM \left[ 1-
\frac{\alpha }{(\alpha ^{3}+R^{3})^{1/3}} \right],
\label{Eddington}
\end{equation}
where the eq. (\ref{gravityforce}) for the force at $v=0$ has been used.

For the used mass of the central object we obtain
$L=6.7 \cdot 10^{39} erg/s$.

The realy luminosity of the central object in the absence of magnetic fields
is of the order of
\begin{equation}
L=v^{2}(r) \stackrel{\cdot}{M} ,
\label{luminosity}
\end{equation}
where  $v(R)$ is defined by eq. (\ref{particlevelocity}). For the
object under consideration $v(R)=2.3 \cdot 10^{8} cm/s$ and at
$\stackrel{\cdot}{M}=10^{-6} M_{\odot}/year$  we obtain
$L= 0.3 \cdot 10^{37} erg/s$.  Thus, in spite of a sufficiently large
the  accretion  rate the object  has a low luminosity.

\section{Ionization Radius}

There must be an ionization zone around the central object with events
horizon which depends on the temperature of the central object and the
physical conditions in the gas environment.

An effective temperature of the object is given by
\begin{equation}
T_{\ast} = \left( \frac{L}{ 4\pi\sigma R^{2} } \right)^{1/4}
\label{Teff}
\end{equation}
where $L$ is the luminosity of the object and
 $\sigma $ is the Stephan-Boltzman constant.

At  $L= 10^{37} erg/s $ the temperature $T_{\ast}=4.8 \cdot 10^{4} K $
. The maximum of the radiation correspons to the vawelength about
$\lambda =500   \stackrel{\circ}{A}$

Let number of neutral atoms , ions and electrons per unit volume be
$n_{1}$ ,  $n_{+}$ and $n_{e}$ , respectively.  We shall assume that
$n_{e}=n_{+}$. At these  conditions $n_{1}=(1-X)n$ , where
$$n=n_{1}+n_{+}$$ is the total number density and
    $$ X=n_{+}/(n_{1}+n_{+})$$  is degree of ionization at the distance
$r$ from the center.

Then at the distance $r$ from thew center the Saha formula  yields:
\begin{equation}
n \frac{X^{2}}{X-1}= B\ W\ \exp(-\tau ),
\label{Saha}
\end{equation}
where
\begin{equation}
B= \frac{g_{+}}{4g_{1}} R^{2} \left( \frac{T_{e}}{T_{\ast}}\right)
 ^{1/2} \frac{2(2 \pi m k T_{\ast})^{3/2}}{h^{3}}
\ln \left( 1 - \exp(\frac{-h \nu_{1}}{k T_{\ast}})\right) ^{-1},
\label{B}
\end{equation}
$T_{\ast}$ is the electronic temperature,  $g_{+}$ and $g_{1}$ are
the statistical weights of the ions and the ground state of the atoms,
respectivily, $\nu_{1}$ is the ionization frequency,
\begin{equation}
W=\frac{1}{2} \left[ 1-\sqrt{1 - \left( \frac{R}{r} \right)^{2}} \right]
\label{dilution}
\end{equation}
is the dilution factor , $\tau$ is the optical thickness that we define as
\begin{equation}
\tau =[1-X(r)]\  k_{\nu } \int_{R}^{r} n(r') dr',
\label{tau}
\end{equation}
where function $X(r)$ is the solution of eqs. (\ref{B}) and
(\ref{tau}) . The constant $k_{\nu }$ is the averaged absorbtion coefficient.

The function $n(r)$ can be found as
\begin{equation}
n(r)=\frac{\stackrel{\cdot}{M}}{4 \pi  r^{2} m_{p} v}
\label{n_of_r}
\end{equation}

and $v(r)$ is the solution of eq.(\ref{eqfor_v}).

For numerical estimates we assume that $T_{e}$ as the function of the
distance $r$ is given by

\begin{equation}
T_{e}=T_{e \infty} \left( \rho / \rho _{\infty} \right)^{\gamma -1}
\label{Temper}
\end{equation}
Setting , for example ,  $T_{\ast}= 6 \cdot 10^{4}\  K $
(which corresponds to $L= 10^{37} \ erg/s $ ) and
$T_{e}= 220\  K  $ (which corresponds to $v_{\infty}=10^{7}\  cm/s $)
we obtaine by a numerical solution of eq. (\ref{rad_ioniz})
the function $X$  of $r$ . The function $Z(r)$ decreases from $1$
to $0.1$ when $r$ increases from $R$ to $10^{14} cm$.
Fig. 3 shows the function of the ionization degree $X$ from the
$z=log_{10}(r)$ at the interval from $R$ to $1\cdot 10^14 \ cm$.

\section{Conclusion}

We have considered the simplest model of the compact object without
events horizon in gas medium with properties of the Galactic center.
It leads to some available for observations   consequences.
. The observation data speak  in favour these consequences rather
than  against them. Consequently, we have an alternative to the supermassive
Black Hole hypothesis. We hope that later a more detail consideration of
the problem  and an analysis of observations will lead to more definite
conclusions.

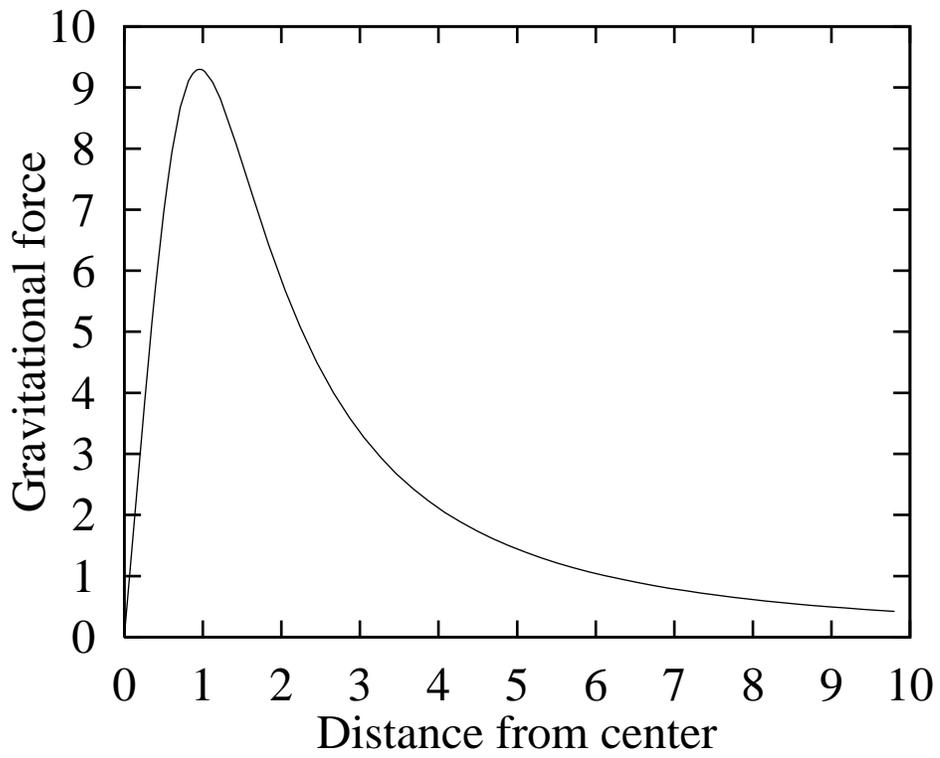
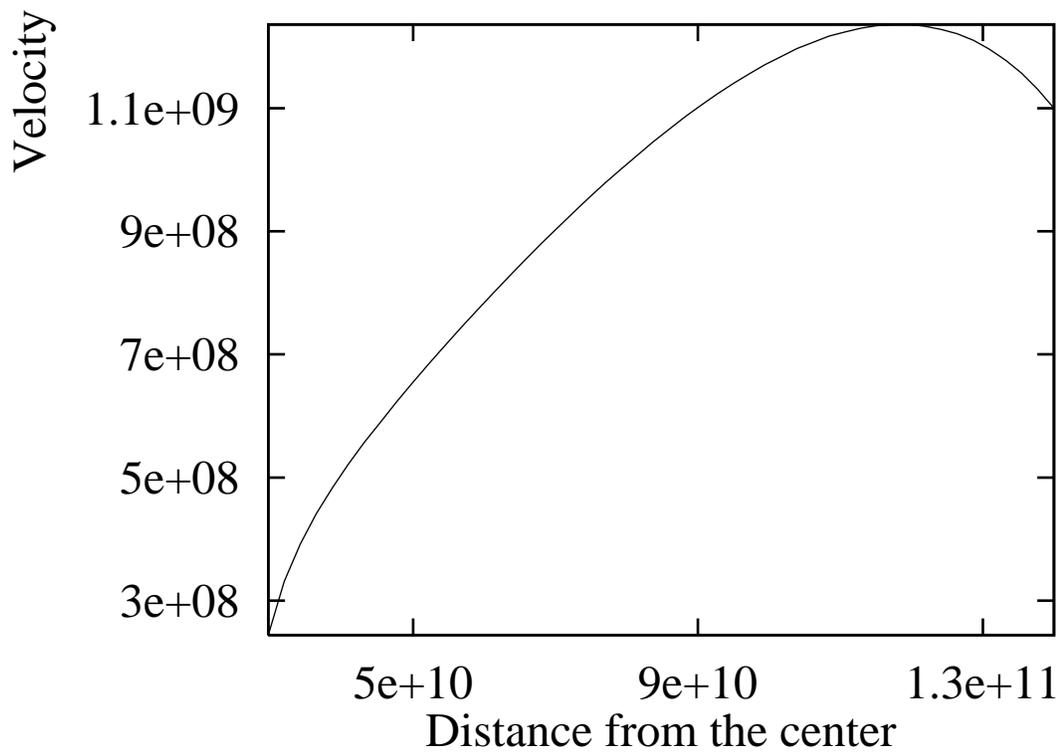
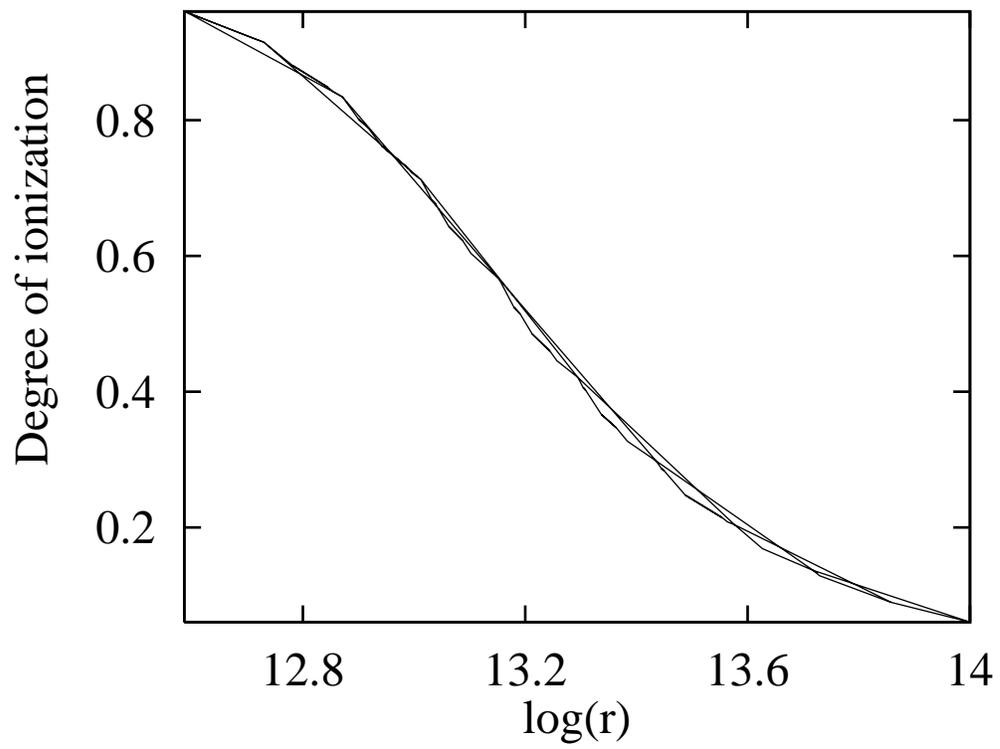
\end{document}